\documentclass[aps,pre,reprint,superscriptaddress,amsmath,showpacs,amssymb]{revtex4-1}
\usepackage{graphicx}
\usepackage{tikz}
\usepackage{amssymb,latexsym}
\usepackage{amsfonts}
\usepackage{color}
\usepackage{epsfig}
\usepackage{graphicx}
\usepackage{pgfplots}
\usepackage{empheq}

\newcommand{\caQ}{{\mathcal Q}}

\newcommand{\mean}[1]{{\left< #1 \right>}}

\begin{document}

\newcommand\mpi{\affiliation{Max Planck Institute for Mathematics in the Sciences, Inselstr. 22, D-04103 Leipzig, Germany}}
\newcommand\uleip{\affiliation{Institut f\"ur Theoretische Physik, Universit\"at Leipzig,  Postfach 100 920, D-04009 Leipzig, Germany}}

\title{Microscopic derivation of the hydrodynamics of active-Brownian-particle suspensions}

\author{Stefano Steffenoni}\email{stefano.steffenoni@mis.mpg.de}\mpi\uleip
\author{Gianmaria Falasco}\email{falasco@itp.uni-leipzig.de}\uleip
\author{Klaus Kroy}\email{klaus.kroy@uni-leipzig.de} \uleip

\date{\today}

\begin{abstract}
We derive the hydrodynamic equations of motion for a fluid of active
particles described by underdamped Langevin equations that reduce to
the Active-Brownian-Particle model, in the overdamped limit. The
contraction into the hydrodynamic description is performed by locally
averaging the particle dynamics with the non-equilibrium many-particle
probability density, whose formal expression is found in the
physically relevant limit of high-friction through a
multiple-time-scale analysis.  This approach permits to identify the
conditions under which self-propulsion can be subsumed into the fluid
stress tensor and thus to define systematically and unambiguously the
local pressure of the active fluid.
\end{abstract}

\date{\today}

\maketitle

\section{Introduction}
The surging field of active matter aims for a microscopic
understanding and control of the material properties of assemblies of
interacting active elements. A main strategy is to revisit the
paradigms of established many-body theories for inanimate matter, and
to elucidate the new physics arising from the nonequilibrium energy
consumption and possibly unusual interactions of its ``atoms''. If
successful, this might help to classify such seemingly diverse systems
as artificial self-propelled colloidal particles \cite{drey05}, motile bacteria
\cite{fri09}, or even flocking birds \cite{bal08}, as examples of a
unified new state of matter, and to inform potential biological or
medical applications \cite{bay15,med15}.

A versatile model for systematic theoretical investigations is
provided by an active-particle suspension.  Despite its simplicity, it
captures the most crucial property of active matter, namely its
intrinsic nonequilibrium. The latter is not brought about by the
application of external forces to a quiescent system, but rather by
the continuous local energy input that fuels the autonomous particle
motion, itself. For experimental realizations, one can draw on a large
arsenal of technologies for tuning the particle-interactions and
propulsion. This promises good control over the various mechanisms by
which the local breaking of detailed balance can manifest itself on
the material level \cite{ste16, fod16, har14, ang11}. For example, if spatial symmetries are broken
(e.g.\ by funnel barriers or asymmetric obstacles), this allows the
microscopic activity to be concerted. Thereby, persistent macroscopic
currents can be induced \cite{cat12} and gears can be set in perpetual
motion \cite{dil10,mal14b}, so that macroscopic work can be extracted
from the microscopic activity \cite{rei16}.  As a consequence, the
status of basic thermodynamic notions, such as temperature and
pressure, on which any coarse-grained description of equilibrium
many-body systems relies, has become a matter of debate even for the
simplest of such model systems \cite{sol15b,tak15}. Only in the limit
of weakly persistent particle motion and weak interactions, the
associated conceptual and practical problems were unambiguously
resolved. An effective Hamiltonian can then be assigned to the
equations of motion, with the activity subsumed into an appropriate
nonequilibrium noise term. This gives rise to an effective
Maxwell-Boltzmann steady state, weighted by an effective temperature,
from which also the pressure can be derived \cite{mar16, mar15,
  fod16}. But this somewhat trivial limit essentially amounts to throwing out the
baby with the bath water, and it is clearly of interest to venture
beyond it and systematically address the more spectacular effects
alluded to above.

For equilibrium systems, pressure can be defined in three ways: (a) as
the derivative of a free energy, (b) as the mechanical force per unit
area on confining walls, and (c) as the trace of the hydrodynamic
stress tensor, which represents the momentum flux in the
system. Definitions (b) and (c) apply even out of equilibrium, where a
unique concept of free energy is still lacking, as they are based on
purely kinematic (mechanical) arguments. Indeed, they have already
been employed within different theoretical approaches: overdamped
Fokker-Planck equations (FPE) \cite{sol15}, the virial theorem for
Langevin equations \cite{win15, fal16}, empirical continuum models
supported by numerical simulations \cite{mar15}, and density functional
theories for overdamped systems \cite{far15}\footnote{\label{Note1}Personal
  Communication by Raphael Wittkowski from Westf{\"a}lische
  Wilhelms-Universit{\"a}t M{\"u}nster}. All these derivations
consider as their starting point the Active Brownian Particle (ABP)
model \cite{sol15, bec16}, in which overdamped particles are allowed
to perform translational and rotational Brownian motion, and the
activity is ascribed to a propulsion force of constant magnitude
along the instantaneous particle orientation. Despite its analytical
and numerical simplicity, this model allows to include many features
of active particles, but it neglects their hydrodynamic
interactions via the solvent flow they excite. Using the nomenclature introduced in Ref.~\cite{mar13}, ABP
is a model of ``dry'' active matter, in contrast with ``wet'' models, in
which the momentum exchanged among the active particles and their
solvent is taken more seriously.

Two main facts are so far agreed upon. First, pressure is, in general,
not a state function, as it depends on the system's microscopic
features and hence cannot be expressed only in terms of thermodynamic
variables.  Secondly, the pressure of an interacting ABP suspension
exhibits a non-monotonic density dependence that manifests itself in a
self-caging or clustering of the particles and a tendency to
accumulate at (curved) walls \cite{yan15,sma15, nik16}. A model-independent,
``thermodynamic'' notion of pressure has so far only been established
for particular cases, such as the mentioned limit of small persistence
of the active motion, assuming non-interacting particles, or at least
torque-free pairwise inter-particle forces, with torque-free wall
interactions. Such special conditions can only (approximately) be
realized for dilute and weakly confined suspensions of spherical
self-propelled particles \cite{sol15, yan15}. The pressure is then
surmised to depend only on bulk properties, which are usually assumed
to be homogeneous and isotropic, and the wall-force is neglected as a
subdominant surface term. For example, Yang et al.\ calculate the
pressure via the Irving-Kirkwood formula \cite{irv50} for the stress
tensor, but consider a spatial average, while Winkler et
al.\ \cite{win15}  obtain the pressure from the virial theorem,
but assume it to be uniform. As a consequence, possible
inhomogeneities necessarily remain hidden, in all cases.

The microscopic derivation of hydrodynamic equations of motion
arguably represents a natural framework to shed light on these issues
and to venture beyond the limitations of current theories.  The recent
literature provides a wealth of hydrodynamic theories derived
phenomenologically, i.e., based on the macroscopic space-time
symmetries \cite{ton95, hat04, ram10}.  A potential difficulty with such
approaches to active matter may be seen in the anticipated breaking of
microscopic symmetries at a mesoscopic level \cite{ste16}, which 
could potentially jeopardize the derivation and
judicious application of the phenomenological hydrodynamic
equations. Yet, attempts to derive them directly from the underlying
microscopic equations of motion are rare. A notable exception is the
work by Bertin et al.\ \cite{ber09}, which is however only valid for
an infinitely dilute gas of active particles with certain specific
alignment interactions. Though exceptionally valuable for the
understanding of the emergence of collective behavior, such approaches
tell us little about an increasing number of interesting experimental
systems characterized by high densities, potentially complex mutual
interactions of the particles, and often narrowly confining
geometries.  Therefore, our aim is to derive the hydrodynamic
equations for an active fluid by a systematic coarse graining of the
(underdamped) microscopic equations of motion of a potentially
strongly interacting and dense active particle suspension.

In the present contribution, we exemplify the procedure for a swimmer model
that neglects the hydrodynamics of the solvent and reduces to the ABP
model in the limit of large friction. We thus only deal with ``dry
swimmers'', here, and defer the discussion of a more realistic
microscopic model to a future contribution. In particular, we derive
the balance equation for the local momentum, which allows us to
uniquely identify the pressure from definition (c). The obtained
hydrodynamic equations keep track of the local inhomogeneities of the
fluid through local averages, performed over the many-body microscopic
probability function. Section $\rm II$ introduces the microscopic model
and, on a formal level, the hydrodynamic equations for the relevant
macroscopic fields, namely, particle density, momentum and
polarization. In sections $\rm III-IV$ we develop a multiple-scale theory
that helps us to close this set of conservation equations based on a
systematic coarse graining of the underlying microscopic model.  In
particular, we apply it to the momentum equation to access the
high-friction limit and obtain a closed expression for the stress
tensor. In section $\rm V$ we investigate the slowest dynamics in the
system, as captured by the equation of motion for the particle
density, neglecting momentum dynamics, in order to make contact with
previous work on the ABP model.  Finally, we derive some explicit
results for the stationary pressure of an ABP
suspension interacting by a hard-sphere repulsion.

\section{Derivation of the hydrodynamic equations} \label{Derivation of the hydrodynamic equations}
Consider the equations for $N$ active Brownian particles in $D=2$ with coordinates $(x_i,v_i, \theta_i) \in \mathbb{R}^{2D+1}$ immersed in a fluid providing friction and noise (unitary masses):
\begin{align}\label{forcedrift}
&\dot x_i = v_i\,\,\,\,\,\,\,\dot \theta_i= \sqrt{2D_{r}}\chi_i \nonumber\\
\dot v_i = -\gamma v_i &+F_i^{\text{int}} +F^{\text{A}}n_i+F_i^{\text{ext}} + \sqrt{2\gamma T} \xi_i
\end{align}
where $F_i^{\text{int}}= \sum_{j \neq i}^N F^{\text{int}}_{ij}$, and $F^{\text{int}}_{ij}$ is the interaction force exerted by particle $j$ on particle $i$, which may depend on both positional and angular coordinates. Possible external forces (e.g. confining walls, gravity) are included in $F_i^{\text{ext}}$. The noises $\xi_i$ and $\chi_i$ are standard Gaussian ones with zero average value and delta correlations in time, $\gamma$ and $T$ are the friction and temperature of the embedding fluid respectively ($k_B=1$). Activity manifests itself through the propulsive force, having magnitude $F^\text{A}=\gamma v_0$ and direction along the particle versor $n_i$, defined by the orientation angle $\theta_i$ and randomized by rotational diffusion at rate $D_r$.

Hydrodynamic fields can be defined from the microscopic dynamics by local ensemble averaging, which for simplicity of notation we denote, for any observable $a_i$ by
\begin{align}\label{<>x}
 \left\langle \sum_i a_i \right  \rangle_{r}\equiv\left\langle \sum_i a_i \delta\left(x_{i}-r\right)\right\rangle.
\end{align} 
The average $\mean{.}$ is taken with respect to the $N$-particle probability density function (PDF)  $\rho_N(\{x_i,v_i,\theta_i\}_{i=1}^N,t)$. The relevant hydrodynamics fields are the fluid mass (or number) density,
\begin{align}
\rho\left(r,t\right)
\equiv\mean{\sum_{i=1}^N 1}_r,
\end{align}
the flow momentum,
\begin{align}
u\left(r,t\right)\rho\left(r,t\right)
\equiv\mean{\sum_{i=1}^N v_i}_r,
\end{align}
and the fluid polarization,
\begin{align}
\mathcal{P}\left(r,t\right)\rho\left(r,t\right)
\equiv\mean{\sum_{i=1}^N n_i}_r.
\end{align}
with $\mathcal{P}\left(r,t\right)\in [0,1]$. For later convenience we introduce here the divergence of the Irving-Kirkwoord (IK) tensor 
\begin{align}
\nabla_{r}\cdot\sigma_{\text{IK}} \equiv\left\langle \sum_{i=1}^{N}F_{i}^{\text{int}}\right\rangle_r,
\end{align}
which can be defined whenever $F^{\text{int}}_{ij}=-F^{\text{int}}_{ji}$ \cite{irv50}, irrespective of the functional dependence of $F^{\text{int}}_{ij}$. The dynamical equations for these observables can be derived from the formula $\partial_t\mean{a}=\mean{La}$, where 
\begin{align}\label{L}
L=\sum_{i=1}^N \bigg[ v_i \cdot \nabla_{x_i} + &(-\gamma v_i + F^{\text{int}}_i +F^{\text{A}}n_i+F_i^{\text{ext}}) \cdot \nabla_{v_i}  \nonumber\\&+\gamma T \nabla^2_{v_i} + D_{r} \partial^2_{\theta_i} \bigg],
\end{align}
is the backward operator associated with \eqref{forcedrift}, for any generic state observable $a$. For the density $\rho\left(r,t\right)$ we find the continuity equation
\begin{align}\label{cont}
\partial_t \rho\left(r,t\right)+\nabla_{r}\cdot (u(r,t)\rho(r,t))=0.
\end{align}
For the polarization density $\mathcal{P}\left(r,t\right)\rho\left(r,t\right)$
\begin{align}\label{Pol}
\partial_t \left(\mathcal{P}(r,t)\rho(r,t)\right) = -\left[D_r\mathcal{P}(r,t)+\nabla_r \cdot\mathcal{C}_{nv}\right]\rho(r,t),
\end{align}
in which we have introduced the correlation tensor
\begin{align}\label{cnv}
\mathcal{C}_{nv}(r,t)\rho(r,t)\equiv \mean{\sum_{i=1}^N n_i v_i }_r.
\end{align}
If the fluid is isotropic and homogeneous, \eqref{cnv} can be factorized into $\mathcal{C}_{nv} \rho= \mathcal{P} u \rho$. However, in many physical situations these symmetries are broken. Examples are cluster formation at high densities \cite{marc16}, and particle accumulation close to boundaries \cite{yan14, wen08}. In these cases $\mathcal{C}_{nv}$ plays an important role, as it keeps track of correlations between particle orientation and velocity. Note that, due to the first term on the $\rm RHS$ of \eqref{Pol}, coming from the rotational diffusion, the polarization locally conserved field. Instead it is locally dissipated at rate $D_r$. For the momentum density $u\left(r,t\right)\rho\left(r,t\right)$
\begin{align}\label{Motionforce}
\partial_t &\left(\rho(r,t)u(r,t)\right)+\nabla_r\cdot\left(\rho(r,t)u(r,t)u(r,t)\right)= \nabla_r\cdot \sigma\nonumber \\+&F^{\text{A}}\mathcal{P}(r,t)\rho(r,t)  -\gamma u\left(r,t\right)\rho(r,t)+F^{\text{ext}}(r)\rho(r,t),
\end{align}
where we identified the full stress tensor $\sigma=\sigma_{\text{IK}}+\sigma_{\text{kin}}$. In the kinetic term, 
\begin{align}\label{sigmaex}
\sigma_{\text{kin}} \equiv-\mean{\sum_{i=1}^N \left(v_i-u\right)\left(v_i-u\right)}_r
\end{align}
advective contributions have been subtracted. According to \eqref{Motionforce}, activity behaves as an external force, on a par with $F_i^{\text{ext}}$ \cite{yan15}. In other words, it is responsible for local violations of momentum conservation and cannot be included in the stress tensor. Yet, if the time derivative of the polarization is negligible, such local non-conservation of momentum can be neglected and the activity can be absorbed in the stress tensor. Indeed, the stationary solution of \eqref{Pol}, $\nabla_r \cdot\left(\mathcal{C}_{nv}\rho\right)=-D_r\mathcal{P}\rho$, renders \eqref{Motionforce} in the form
\begin{align}\label{Motionforce2}
\rho\frac{du}{dt}= \nabla_r\cdot\sigma_{\rm s}-\gamma u \rho +F^{\text{ext}}\rho
\end{align}
with
\begin{align}\label{sigma}
\sigma_{\rm s} \equiv \sigma_{\text{IK}}+\sigma_{\text{kin}}-\frac{F^\text{A}}{D_r}\mathcal{C}_{nv}\rho
\end{align}
We note that on the $\rm LHS$ of Eq. \eqref{Motionforce2} we have introduced the material derivative, as usually done. Since the first summand on the $\rm RHS$ of \eqref{Pol} is a constant damping term, the condition $\partial_t\left( \mathcal{P}\rho\right)=0$ may be guaranteed by a fast relaxation of the particle orientation in comparison with the shortest  accessible time scale $dt$, i.e. $D_{r} dt \gg 1$. 
Large $D_r$ corresponds to particles experiencing strong rotational diffusion, so that the persistence length of their paths is small in comparison with the mean particle distance and the characteristic system length and the only relevant effect of activity is an enhanced translational diffusion \cite{lau11}. Formally, the system then behaves as a passive Brownian particle suspension maintained at an elevated effective temperature \cite{tak15, yan14, pal10}. 

In order to make the hydrodynamic description self-contained, we finally have to express the microscopic degrees of freedom, in terms of the hydrodynamic variables. To this end, we derive a suitable expression for the $N$-particle PDF valid in the limit of high-friction, in the next section.

\section{Multiple time scale theory}
With typical suspensions of active colloidal particles in mind, we now consider the limit of large friction $\gamma$, where the particle velocity $v_i$ relaxes rapidly in comparison with the position $x_i$. We can then treat the latter as adiabatically slow as compared to the former. The scale separation can formally be exploited by means of a multiple time scale theory \cite{hol12}. With the aim of systematically performing the overdamped limit for the microscopic many-body PDF, we introduce a dimensionless small parameter $\epsilon \ll 1$, such that $\gamma=\epsilon^{-1}\tilde{\gamma}$ with $\tilde{\gamma}=O(1)$. This permits the distinction of three time scales, namely, fast $\tau_0=\epsilon^{-1}t$, intermediate $\tau_1=t$, and slow $\tau_2=\epsilon t$, and to expand the $N$-particle PDF $\rho_{N}(\left\{x_{i},v_{i},\theta_{i}\right\}^{N}_{i=1}, t)$ associated with \eqref{forcedrift} as
\begin{align}\label{rho_exp}
\rho_{N}=\rho^{(0)}_{N}+\epsilon \rho^{(1)}_{N}+ O(\epsilon^2).
\end{align}
In the absence of activity such multi-scale analysis  leads to the overdamped FPE, or Smoluchowski equation, for a passive $N$-particle system. A standard derivation can be found, e.g., in Ref.~\cite{boc97}. As a byproduct, it yields a formal expression for $\rho_{N}$. While the Maxwell-Boltzmann distribution is \textit{a priori} known to provide the stationary state for an equilibrium system, an explicit calculation is required for active and driven systems \cite{bra07}

Our starting point is the FPE derivable from \eqref{L},
\begin{align}\label{FPE}
\frac{\partial\rho_{N}}{\partial t}+&\sum_{i=1}^N \bigg[ \left(v_{i}\cdot\nabla_{x_{i}}+\left(F_i^{\text{ext}}+F_{i}^{\text{int}}\right)\cdot\nabla_{v_{i}}\right)-D_{r}\partial^2_{\theta_i} \bigg]\rho_{N} \nonumber \\ &=\gamma\sum_{i=1}^N\nabla_{v_{i}}\cdot\left(v_{i}-v_0n_i+T\nabla_{v_{i}}\right)\rho_{N},
\end{align}
in which the active force $\gamma v_0 n_i$ is seen to act like a negative friction along the orientation direction. Therefore, it contributes to the velocity relaxation, implying that the velocity PDF will depend on the orientation angle, differently from a passive system. 
Following Ref. \cite{boc97}, we derive the density $\rho_{N}$ to the order $\epsilon$ in Appendix \ref{app}. We merely summarize its results, here. \\
Both terms in \eqref{rho_exp} involve a Gaussian weight for the correlations of velocities and orientations. Accordingly, we find the zero-order density 
\begin{align}\label{rho0}
\rho^{(0)}_{N}=  \Phi\frac{1}{2\pi T}\exp{\left(-\frac{(v_i-v_0n_i)^2}{2T}\right)},
\end{align}
and the first correction 
\begin{align}\label{rho1}
\rho^{(1)}_{N}= \frac{1}{\tilde{\gamma}}&\left[\Psi-\left(v_i-v_0n_i\right)\left(\frac{\partial\Phi}{\partial x_i}-\frac{\left(F^{\text{int}}_{i}+F_i^{\text{ext}}\right)}{T}\Phi\right)\right]\nonumber\\&\cdot\frac{1}{2\pi T}\exp{\left(-\frac{(v_i-v_0n_i)^2}{2T}\right)},
\end{align}
where $\Phi(\left\{x_{i},\theta_{i}\right\}^{N}_{i=1},  \tau_1, \tau_2)$ and $\Psi(\left\{x_{i},\theta_{i}\right\}^{N}_{i=1}, \tau_1, \tau_2)$ are unknown $N$-particle functions. Ignoring higher order contributions, they give the overdamped $N$-particle PDF
\begin{align}\label{rhover}
\int \prod_{i=1}^N dv_i \rho_{N} = \Phi+ \frac{\epsilon}{\tilde{\gamma}} \Psi+\mathcal{O}(\epsilon^2).
\end{align}
By construction, this PDF satisfies the overdamped FPE  \cite{sol15b}, which can also directly be derived by ignoring the particle inertia in \eqref{forcedrift}. 

\section{Hydrodynamic equations in the high-friction limit}

\subsection{Momentum Equation}

In the previous section we have seen that the large friction experienced by the particles allows for an expansion \eqref{rho_exp} of the $N$-particle PDF in powers of the friction $\gamma$. In the present section, we consistently make use of \eqref{rho_exp}, together with \eqref{rho0} and \eqref{rho1}, in order to expand the tensor
\begin{align}\label{sigkin_e}
\sigma_{\rm kin}=\sigma_{\rm kin}^{(0)}+\epsilon\, \sigma_{\rm kin}^{(1)}+\mathcal{O}(\epsilon^2).
\end{align}
The expansion of $\sigma_{\rm kin}$ utilizes the notation of \eqref{rho_exp}, e.g. $\sigma_{\rm kin}^{(0)}$ involves only averaging over \eqref{rho0}, namely,
\begin{align}\label{kindom}
-\sigma_{\rm kin}^{(0)}&=\sum_{i=1}^{N}\int dx_{i}d\theta_{i}\,\left[T \mathbb{I}+(v_0n_i-u)^2\right]\delta\left(x_{i}-r\right)\Phi\nonumber \\
&= T\rho^{(0)}\mathbb{I}+\left\langle \sum_{i=1}^{N}\left(v_{0}n_{i}-u\right)^{2}\right\rangle_r^{(0)}
\end{align}

Here, $ \mathbb{I}$ is the identity matrix and $\mean{\dots}^{(0)}$ denotes an average performed with $\rho^{(0)}_{N}$, as given by \eqref{rho0}, e.g. $\rho^{(0)}=\bigl\langle\sum_{i=1}^N 1\bigr\rangle^{(0)}_r$. From \eqref{kindom}, it is clear that the active contribution prevents the velocity fluctuations from relaxing to the heat bath temperature on the fast time scale. The second-leading term, denoted by $\mean{\dots}^{(1)}$, is found by using \eqref{rho1}: 
\begin{align}\label{vv1}
-\sigma_{\rm kin}^{(1)}&=\sum_{i=1}^{N}\int dx_{i}d\theta_{i}\,\left[T \mathbb{I}+(v_0n_i-u)^2\right]\delta\left(x_{i}-r\right)\frac{\Psi}{\tilde{\gamma}} \nonumber \\+ \frac{2}{\tilde{\gamma}}\int &\prod_{i=1}^{N}dx_{i}d\theta_{i}\,\sum_{i=1}^{N}\delta\left(x_{i}-r\right)(v_0n_i-u)\left(F^{\text{int}}_{i}+F_i^{\text{ext}}\right)\Phi  \nonumber \\- \frac{2T}{\tilde{\gamma}}&\int \prod_{i=1}^{N} dx_{i}d\theta_{i}\,\sum_{i=1}^{N}\delta\left(x_{i}-r\right)(v_0n_i-u)\frac{\partial\Phi}{\partial x_{i}}.
\end{align}
With slight abuse of notation, here we denote by $\Psi$ and $\Phi$ the functions in \eqref{rho0} and \eqref{rho1} integrated over the coordinates of the $N-1$ particles $j \neq i$. According to \eqref{rhover}, we identify the first term in the integral in \eqref{vv1} with  $T\rho^{(1)} \mathbb{I}$. Integrating by parts in the third line and using the definition of $\rho^{(0)}$ we get
\begin{align}\label{vv1_2}
-\sigma_{\rm kin}^{(1)}&=T\rho^{(1)}\mathbb{I} +\left\langle \sum_{i=1}^{N}\left(v_{0}n_{i}-u\right)^{2}\right\rangle _{r}^{(1)}\nonumber\\+\frac{2}{\tilde{\gamma}}&\left\langle\sum_{i=1}^{N}\left(F^{\text{int}}_i+F_i^{\text{ext}}\right)(v_0n_i-u)\right\rangle^{(0)}_r\nonumber \\&-\frac{2T}{\tilde{\gamma}}\nabla_{r}\left\langle \sum_{i=1}^{N}(v_0n_i-u)\right\rangle^{(0)}_r .
\end{align}
The first line extends the leading contribution in \eqref{kindom} to the next order. The remaining terms constitute the non-equilibrium corrections to the kinetic tensor and they will be considered in the following. Reverting to the physical quantity $\gamma$, the sum of the kinetic tensors gives
\begin{align}\label{sigmakin2}
-\sigma_{\rm kin}^{(0)}&-\epsilon\, \sigma_{\rm kin}^{(1)} \simeq T\rho +\left\langle \sum_{i=1}^{N}\left(v_{0}n_{i}-u\right)^{2}\right\rangle_r
 \nonumber \\+\frac{2}{\gamma}&\left\langle \sum_{i=1}^{N}\left(v_{0}n_{i}-u\right)\left(F_{i}^{\rm ext}+F_{i}^{\rm int}\right)\right\rangle_r \nonumber \\&-\frac{2T}{\gamma}\nabla_{r}\left\langle \sum_{i=1}^{N}\left(v_{0}n_{i}-u\right)\right\rangle_r
 \end{align}
 All the average values in \eqref{sigmakin2} contain the hydrodynamic flow velocity $u$. In order to obtain a consistent expansion in powers of the friction $\gamma$, $u$ must itself be expanded to order $\mathcal{O}(1/\gamma)$, namely
\begin{align}\label{uexp}
\rho u=v_{0}\mathcal{P}\rho-\frac{T}{\gamma}\nabla_{r}\rho+\frac{1}{\gamma}\left(\nabla_{r}\cdot\sigma_{\rm IK}+F^{\rm ext}\rho\right)
 \end{align}
The physical interpretation is that the ${O}(\gamma^0)$ coherent velocity is an active streaming contribution $v_0\mathcal{P}$, while the usual hydrodynamic streaming terms are damped by factor $\gamma^{-1}$ term in the microscopic model, Eq. \eqref{forcedrift}. Plugging \eqref{uexp} into \eqref{sigmakin2}, we obtain for the kinetic tensor \eqref{sigkin_e}
\begin{align}\label{sigmakin3}
 -\sigma_{\rm kin}&=T\rho+v_{0}^{2}\left(\mathcal{Q}-\mathcal{PP}\right)\rho-\frac{2v_{0}\mathcal{P}}{\gamma}\left(\nabla_{r}\cdot\sigma_{\rm IK}+F^{\rm ext}\rho\right)
 \nonumber \\&+\frac{2v_{0}}{\gamma}\left\langle \sum_{i=1}^{N}\left(F_{i}^{\rm ext}+F_{i}^{\rm int}\right)n_{i}\right\rangle_r +\mathcal{O}(1/\gamma^{2})
 \end{align}
where defined
\begin{align}\label{nematic}
 \mathcal{Q}\left(r,t\right)\rho(r,t)\equiv\left\langle \sum_{i=1}^{N}n_{i}n_{i}\right\rangle_r,
\end{align}
in line with what we said about the order of the terms in \eqref{uexp} the ${O}(\gamma^0)$ correction to the equilibrium contribution to $\sigma_{\rm kin}$ is due to active noise, while the ${O}(1/\gamma)$ terms subtract the usual streaming contribution from it.  
If $r$ is a point far from a boundary, and the density is low enough to prevent cluster or lanes formation \cite{men13}, both $\mathcal{P}$ and the off-diagonal components of $\mathcal{Q}(r,t)$ are zero and the active noise term becomes $v_0^2\mathbb{I}\rho$ divided by the space dimension. \\ 
Together with the ordinary kinetic pressure it may then be identified as arising from an effective temperature. More generally, though, the temperature of an active particle system is, in principle, neither homogeneous nor isotropic. Nevertheless,  since the entries of $\mathcal{Q}$ and $\mathcal{P}$ are bounded by one and, in typical experiments, $v_0 \lesssim 100 \,\mu \text{m/s}$, the active correction is much smaller than the thermal energy and can thus be neglected in practice. In other words, the heating of an active particle system due to the activity itself (if not by its propulsion machinery) is usually minute.\\
The remaining terms constitute the non-equilibrium corrections to the kinetic tensor, which evidently depend on the system's microscopic details. The correlation between $F^{\text{int}}_{i}$ and $n_{i}$ prevents an interpretation following the standard $\rm IK$-tensor derivation. Indeed, due to activity, it is not possible to extract a gradient with respect to $r$ from the expression in the second line of \eqref{vv1_2}. Likewise, the external force contribution is not generally factorizable. We therefore define the local tensors
\begin{align}
\mathcal{I}_1 \left(r,t\right)\rho(r,t)&=\left\langle \sum_{i=1}^{N}F_i^{\text{ext}}n_{i}\right\rangle_r,\label{I1}\\
\mathcal{I}_2 \left(r,t\right)\rho(r,t)&=\left\langle \sum_{i=1}^{N}F^{\text{int}}_{i}n_{i}\right\rangle_r\label{I2},
\end{align}
that account for the correlations between the particle orientation and the external and internal forces, respectively. We recall that $F^{\rm ext}$ is considered to be a generic external force. If it represents the interaction with a wall, the term $\mathcal{I}_1(r,t)\rho(r,t)$ describes the correlation among the wall force and the local polarization. In many previous works \cite{sol15b, win15, spe16} the integral of this term over the whole fluid is ignored since it only contributes a subdominant surface term. This approximation is valid when the persistence length is much smaller than the characteristic length of the system, and it becomes exact in the thermodynamic limit. 
However, when long-range correlations are induced by the external forces, or when $\mathcal{I}_1$ is considered locally near a wall or an obstacle, it is not negligible and may be responsible for such effects as wall accumulation or ratcheting. If relevant, these effects undermine attempts to interpret the pressure as a state function in the conventional (broad) sense \cite{sol15}. \\

Summing up, the balance equation for the momentum of the active fluid becomes
\begin{align}\label{uover}
\rho \frac{du}{dt}=\nabla_r\cdot \sigma-\gamma\rho u+\gamma v_0\mathcal{P}\rho+F^{\text{ext}}\rho,
\end{align}
with the stress tensor to order $\mathcal{O}(1/\gamma)$ given by 
\begin{align}\label{sigmaover}
\sigma=&-T\rho \mathbb{I} +\sigma_{\text{IK}}-\frac{2v_{0}}{\gamma}\left(\mathcal{I}_{1}+\mathcal{I}_{2}\right)\rho\nonumber \\&+\frac{2v_{0}}{\gamma}\mathcal{P}\left(\nabla_{r}\cdot\sigma_{\rm IK}+F^{\rm ext}\rho\right).
\end{align}
Its negative trace, normalized by the space dimension, defines the local fluid pressure \cite{irv50}. The second line can be understood as a nonequilibrium streaming contribution subtracted from the active stresses in the first line. 
Considering the limit $\gamma \to \infty$, \eqref{sigmaover} validates (and extends to interacting particles) a result by Speck and Jack \cite{spe16}, namely that pressure is independent of activity. Only for noninteracting particles in a homogeneous phase away from any boundaries the result remains valid even to $\mathcal{O}(1/\gamma)$.
Under more general conditions, the pressure clearly differs from that of an equilibrium fluid. This should be expected, since static properties of a nonequilibrium system are known to depend on its dynamical parameters \cite{zia07}.

\subsection{Density equation}\label{Density equation}
From \eqref{uover} we can derive a dynamical density equation for the number or mass density $\rho(r,t)$, making contact with previous works on density functional (field) theory for overdamped active particles \cite{Note1}. Neglecting the time derivative of the velocity in \eqref{uover} in the case of steady flow, we find an expression for the stationary fluid velocity,
\begin{align}
\gamma u\rho=\nabla_{r}\cdot\sigma+\gamma v_0\mathcal{P}\rho+F^{\text{ext}}\rho
\end{align}
with which we can simplify the continuity equation \eqref{cont} and the equation for the polarization \eqref{Pol}:
\begin{align*}
\partial_{t}\rho+\nabla_r\cdot J_{\rho}=0 \hspace{0.8cm} \partial_{t}\left(\mathcal{P}\rho\right)+\nabla_r\cdot\left(\mathcal{C}_{nv}\rho\right)=-D_{r}\mathcal{P}\rho.
\end{align*}
Here we have introduced the particle flux
\begin{align}
J_{\rho}=\frac{1}{\gamma}\left(\nabla_{r}\cdot\sigma+F^{\text{ext}}\rho\right)+v_0\mathcal{P}\rho, 
\end{align}
and the polarization flux, identified with the tensor $\mathcal{C}_{nv}\rho$. Similar equations have already been presented in Ref. \cite{marc16, sol15}. The former started from the FPE associated to the overdamped version of \eqref{forcedrift}, and the latter considered phenomenological equations supported by numerical simulations. For consistence also $\mathcal{C}_{nv}$ must be expanded in powers of $\gamma^{-1}$. Employing the very same procedure used above for $\sigma_{\text{kin}}$, an expression valid to order $\mathcal{O}(1/\gamma)$ is found,
\begin{align}\label{cnv_fin}
\mathcal{C}_{nv}\rho=v_0\mathcal{Q}\rho+\frac{1}{\gamma}\left(\mathcal{I}_1\rho +\mathcal{I}_2\rho-T\nabla_{r} \left(\mathcal{P}\rho\right)\right).
\end{align}
As pointed out in Sec.  \ref{Derivation of the hydrodynamic equations}, we can neglect the time variation of the polarization in order to simplify $J_{\rho}$. Namely, setting $\partial_{t}\left(\mathcal{P}\rho\right)=0$, we can replace the last term in $J_\rho$ with $ \mathcal{P}\rho = - \nabla_{r}\cdot\left(\mathcal{C}_{nv}\rho\right)/D_{r}$, and obtain 
\begin{align}
J_{\rho}=\frac{1}{\gamma}\nabla_{r}\cdot\left[\sigma-\frac{\gamma v_0^2}{D_{r}}\mathcal{Q}\rho+\frac{v_0T}{D_{r}}\nabla_{r}\left(\mathcal{P}\rho\right)\right]\nonumber \\ -\frac{1}{\gamma}\nabla_{r}\cdot\left[\frac{v_0}{D_{r}}\mathcal{I}_1\rho+\frac{v_0}{D_{r}}\mathcal{I}_2\rho\right]+\frac{1}{\gamma}F^{\text{ext}}\rho \,.
\end{align}
Now we substitute the stress tensor \eqref{sigmaover} and neglect terms $\mathcal{O}\left(1/\gamma^2\right)$
\begin{align}\label{current2}
J_{\rho}=\frac{1}{\gamma}&\nabla_{r}\cdot\left[-T \rho \mathbb{I}+\sigma_{\text{IK}}-\frac{\gamma v_0^2}{D_{r}}\mathcal{Q}\rho+\frac{v_0T}{D_{r}}\nabla_{r}\left(\mathcal{P}\rho\right)\right]\nonumber\\ &-\frac{1}{\gamma}\nabla_{r}\cdot\left[\frac{v_0}{ D_{r}}\mathcal{I}_1\rho+\frac{v_0}{ D_{r}}\mathcal{I}_2\rho\right]+\frac{1}{\gamma}F^{\text{ext}}\rho
\end{align}
For vanishing particle flux, the external force is balanced by the terms included in the divergence, inducing the definition of a stress tensor $\sigma_{\rm s}$ reliable in the regime of stationary polarization field, as pointed out in \eqref{sigma}. \\
\begin{align}\label{current3}
J_{\rho}=&\frac{1}{\gamma}\nabla_{r}\cdot\sigma_{\rm s}+\frac{1}{\gamma}F^{\text{ext}}\rho
\end{align}
Finally, we turn \eqref{current3} into a sum of a purely diffusive flux plus a drift term arising from external forces, $J_{\rho}=-D_{\rm eff}\nabla_{r}\rho+ \frac{1}{\gamma}F^{\text{ext}}\rho$.
This is done, like in equilibrium, relating the gradient diffusion matrix $D_{\rm eff}$ to the compressibility of the fluid. 
Using the chain rule, we can write
\begin{align}
\frac 1 \gamma \nabla \cdot \sigma_{\rm s}= \frac 1 \gamma  \partial_{\rho}  \sigma_{\rm s} \cdot \nabla \rho
\end{align}
and then define the diffusion matrix $D_{\rm eff}$ by introducing the compressibility coefficients matrix $\kappa_T^{ij}$ ($i,j=1,2$),
\begin{align}\label{diffcompr}
(D_{\rm eff})_{ij} \equiv \frac{1}{\gamma \rho} (\kappa_T^{-1})_{ij}, \qquad \kappa_{T}^{ij}\equiv -\frac{1}{\rho}\left(\frac{\partial(\sigma_{\rm s})_{ij} }{\partial\rho}\right)^{-1}_{T}.
\end{align}
The dependence of the stress tensor $\sigma_{\rm s}$ on density $\rho$ is not straightforward, but an expansion in powers of $\rho$ can be performed in a weakly inhomogeneous approximation as presented in Sec. \ref{Weakly inhomogeneous density approximation}.

\subsection{Local fluid pressure and wall pressure}
In the section \ref{Density equation} we have introduced the static stress tensor $\sigma_{\rm s}$. Equivalently, it can be derived from \eqref{Motionforce2} by applying multiple-time-scale theory to \eqref{sigma}, i.e. involving the dynamics characterized by a fast polarization relaxation. Indeed, neglecting the time variation of the polarization, the active contribution to the momentum balance is naturally included in the stress tensor through the correlating tensor $\mathcal{C}_{nv}$ that, up to order $\mathcal{O}(1/\gamma)$, is given by \eqref{cnv_fin}. If plugged into \eqref{sigma}
\begin{align}\label{sigmaneg}
\sigma_{\rm s}=-&T\rho \mathbb{I}+\sigma_{\text{IK}}-\frac{\gamma v_0^2}{D_{r}}\mathcal{Q}\rho
+\frac{v_0T}{D_r}\nabla_{r}(\mathcal{P}\rho)\nonumber \\-&\frac{2v_0}{\gamma}\left(1+\frac{\gamma}{2D_{r}}\right)\rho\left(\mathcal{I}_{1}+\mathcal{I}_2\right)\nonumber\\&+\frac{2v_0}{\gamma}\mathcal{P}\left(\nabla_r\cdot\sigma_{\rm IK}+F^{\rm ext}\rho\right)
\end{align}
Note that some active contributions appearing in \eqref{sigmaover} now have a renormalized pre-factor. Neglecting sub-leading terms in the high-friction limit $\gamma \gg D_r$, 
\begin{align}\label{sigmaneg1}
&\sigma_{\rm s}=-T\rho\mathbb{I}+\sigma_{\text{IK}}-\frac{\gamma v_0^2}{D_r}\mathcal{Q}\rho\nonumber \\+&\frac{v_0}{D_r}\left[T\nabla_{r}(\mathcal{P}\rho)-\rho\left(\mathcal{I}_{1}+\mathcal{I}_2\right)\right].
\end{align}
We stress that the tensors in the last line of \eqref{sigmaneg} are equal to the ones in the second in the absence of correlations. So they are expected to be of the same order. We recall that the appended subscript ``s'' emphasizes that \eqref{sigmaneg1} is valid only under the stationary condition $\partial_t\left(\mathcal{P}\rho\right) \simeq 0$. In turn, the latter condition implies $\partial_t\left(u\rho\right)=\mathcal{O}(1/\gamma)$, for consistency with the assumption $\gamma \gg D_r$. Therefore, $\sigma_{\rm s}$ is best suited to inspect the mechanical equilibrium of an active fluid, as it yields for static conditions ($u=0$) the momentum balance \eqref{Motionforce2} in the form
  \begin{align}\label{sigma_s}
 F^{\text{ext}}\rho= -\nabla_r \cdot \sigma_{\rm s} .
  \end{align}
When $F^{\text{ext}}$ is a confining wall force,  $\sigma_{\rm s}$ gives the local force per unite area exerted by the active fluid on its container. It clearly differs from the local fluid pressure in \eqref{sigmaover}, because the active force is already included in \eqref{sigmaneg} and hence does not  explicitly show up in \eqref{sigma_s}. This becomes apparent when comparing \eqref{sigma_s} with the static limit of \eqref{uover}, 
\begin{align}\label{sigmau=0}
F^{\text{ext}}\rho+\gamma v_0\mathcal{P}\rho= -\nabla_r\cdot \sigma.
\end{align}
It manifestly shows that the external forces $F^{\rm ext}$ are counterbalanced not only by the active pressure $\sigma$ alone, but also by the \emph{internal} body force $\gamma v_0\mathcal{P}$ \cite{spe16}, legitimated to dissipate momentum, only under non-stationary conditions. 

%
In the light of \eqref{sigma_s}, we can define the local pressure exerted on a wall by taking the trace of \eqref{sigmaneg1}. By construction $\text{Tr}\,\caQ=1$, while the trace of the correlating tensors $\mathcal{I}_{1}$ and $\mathcal{I}_2$ boils down to the scalar product of the forces $F^{\text{int}}_i$ and $F_i^{\text{ext}}$ with the orientational angle $n_i$:
\begin{align}\label{Pressure}
&P^{\rm wall}(r)=\rho T-\frac{1}2\text{Tr}\, \sigma_{\text{IK}}+ \nonumber \\ \rho&\frac{\gamma v_0}{2D_r}v\left(\rho(r,t)\right)+\frac{v_0T}{2D_r}\nabla_{r}\cdot(\mathcal{P}\rho).
\end{align}
This expression is consistent with the pressure derived by some of us in \cite{fal16} through the virial theorem. The additional term in \eqref{Pressure} depending on the divergence of the polarization, does not appear in \cite{fal16} since therein only the average pressure on a fluid container is considered, which requires an integration over the whole space. We have defined the density-dependent actual swim speed
\begin{align}\label{veldep}
v\left(\rho(r,t)\right) \equiv v_0+\frac{1}{\gamma} \text{Tr}\, \left(\mathcal{I}_1\left(r,t\right)+\mathcal{I}_2\left(r,t\right)\right),
\end{align}
in which the bare swim speed $v_0$ is renormalized by the effect of inter-particles interactions and external forces. In absence of external forces \eqref{veldep} boils down to the expression by Marchetti et al. \cite{marc16}. Numerical simulations \cite{tak14, mar15} show that, at high density, the effective velocity of active particles is reduced due to self-caging effects, implying that the second term of \eqref{veldep} turns negative. Intuitively, this happens when the particles are swimming oppositely to the mutual force they experience (regardless of the particular form of $F^{\rm int}$), and thus corresponds to trapped configurations. 
Hence, for hard-core interactions, the particle velocities point towards each other in the trapped state, but attractive interactions also allow for trapped configurations with outwards pointing velocities at small activities. 
%
%
%
%

\section{Weakly inhomogeneous density approximation}\label{Weakly inhomogeneous density approximation}
Self-caging phenomena, as just described, are responsible also for the non-monotonic behavior of the pressure as function of density. This becomes evident considering a gradient expansion of \eqref{Pressure}, in which the explicit density dependence, of the tensors $\sigma_{\text{IK}}, \,\mathcal{I}_1, \,\mathcal{I}_2$ is approximately resolved. Namely, the hydrodynamic fields are supposed to change slowly in space, so that gradient terms can be discarded. For simplicity, we suppose that the only external force is a hard-wall potential, and that inter-particle forces $F_{ij}^{\rm int}=-\nabla_i U(|x_i-x_j|)$ derive from a central pair potential. This assumption allows to neglect any term involving $F^{\text{ext}}$, since it amounts to a subleading contribution in the system size when the average pressure is considered.  

We first consider the Irving-Kirkwood tensor, that can be written introducing the pair distribution function $g_2(r,R)$ which measures the spatial correlations between particles at position $r$ and $r+R$~\cite{irv50},
\begin{align}\label{IKcomp}
\sigma_{\text{IK}} =\frac{\rho\left(r\right)}2\int dR\,\frac{R R}{|R|}U'\left(\left|R\right|\right)\rho\left(r+R\right)g_2\left(r,R\right).
\end{align}
The zero order in a gradient-expansion approximation, i.e., $\rho\left(r+R\right)\simeq\rho\left(r\right)$, yields $
\sigma_{\text{IK}}\simeq-\rho^2 a$ with a slowly varying matrix
\begin{align}\label{acomp}
a\left(r\right)=-\frac{1}2\int dR\, \frac{RR}{|R|}U'\left(\left|R\right|\right)g_2\left(r,R\right).
\end{align}
with $\int d \theta g_2(r,\theta,R)= g_2(r,R)$. We proceed similarly for the remaining terms. Using the definition \eqref{I2}, average in $\mathcal{I}_2$, reads explicitly
\begin{align}
\mathcal{I}_{2}\rho&=\sum_{i=1}^{N}\sum_{j\neq i}^{N}\int dx_{i} dx_{j} d\theta_{i}\delta\left(x_{i}-r\right)\rho\left(x_{i},\theta_{i}\right) \nonumber \\  \quad\quad&\qquad \times  n_{i} F_{ij}^{int}(x_i-x_j)\rho\left(x_{j}\right)g_{2}\left(x_{i},\theta_{i},x_{j}\right) \nonumber \\
&=  \int dR d \theta \, n(\theta) F^{\rm int}(R) \rho(r+R)\rho(r,\theta)g_2(r,\theta,R) \label{bcomp}
\end{align}
If the symmetries of the system allow it, the density can be  factorized $\rho(r,\theta)\simeq \rho(r) f(\theta)$,
\begin{align}
\mathcal{I}_{2}\rho= \rho(r)\int dR d \theta \, n(\theta) F^{\rm int}(R) \rho(r+R)f(\theta)g_2(r,\theta,R) \label{bcomp}
\end{align}Hence, to a first approximation we obtain $\mathcal{I}_2 \rho \simeq\rho^2 b$, where 
\begin{align}\label{bcomp}
b(r)=\int dR d \theta \,n(\theta) F^{\rm int}(R) f(\theta)g_2(r,\theta,R),
\end{align}
If we plug \eqref{acomp} and \label{bcomp} into \eqref{Pressure} after taking the trace, we  obtain
\begin{align}\label{Vir}
P^{\rm wall}=\left(T+\frac{\gamma v_0^2}{2D_r}\right)\rho+\frac{1}{2}{\rm Tr}\left(a+\frac{v_0}{D_r}b\right)\rho^2,
\end{align}
Here $g_2(r,\theta,R)$ is the probability of finding a particle in $r$ with orientation $\theta$ and a particle in $r+R$. With slight abuse of notation we use the same symbol to denote the pair density distribution $g_2(r,R)\equiv \int d \theta g_2(r,R,\theta)$. Equation ~\eqref{Vir} is similar to the equilibrium virial equation of state, suggesting the definition of the effective temperature,  
\begin{align}
T_{\rm eff}=T+\frac{\gamma v_0^2}{2D_r}.
\end{align}
It exhibits the ideal-gas contribution, with the usual temperature renormalization due to the stochastic active motion  \cite{tak15, yan14, pal10}. The pressure arising from the interactions has a contribution ${\rm Tr}a$ of the standard form known from equilibrium. It knows about the activity only through the nonequilibrium pair distribution $g_2$. The second term displays an explicit dependence on $v_0$ indicating its absence in equilibrium. The factor ${\rm Tr}b$ can turn negative for self-trapping configurations, and leads to the aforementioned decrease of the actual velocity \eqref{veldep}. Here we have found that the same phenomenon may cause a non-monotonic behavior of the pressure for large enough values of the persistence length  $\ell_p\equiv v_0D_r^{-1}$ compared to the mean particle distance $\rho^{-\frac{1}{2}}$ \cite{yan15}. Similar conclusions have been drawn by Takatori et al. 	from the density expansion of the swim pressure  \cite{tak14} starting from microrheology results.\\\\
Expansion \eqref{Vir} for $P^{\rm wall}$, and equivalently for $\sigma_{\rm s}$ allows a simplification of the compressibility coefficient \eqref{diffcompr} introduced in the diffusion equation. Namely, 

\begin{align}
\frac{1}{\rho}(\kappa_T^{-1})_{ij}=T+\frac{\gamma v_{0}^{2}}{2D_{r}}+\rho\left((a)_{ij}+\frac{v_{0}}{D_{r}}(b)_{ij}\right).
\end{align}

Generally, $(a)_{ij}>0$ for systems with repulsive inter-particle potentials. Hence, at equilibrium we expect them to posses a reduced compressibility compared with an ideal gas. Conversely, since $(b)_{ij}$ can be negative for active particles thanks to their self-caging properties, the compressibility can be increased even for purely repulsive inter-particle forces. In view of \eqref{diffcompr}, an increasing compressibility entails a decreasing diffusivity, as usual. \\\\
%

\section{Conclusion}
For a ``dry" active system, we have derived balance equations for hydrodynamic observables starting from the underlying microscopic dynamics. Particularly, the equation for momentum balance univocally leads to a definition of the pressure via the stress tensor. Our expressions derived in the high-friction limit rationalize the features observed in active systems: violation of (local) momentum conservation due to the active swim force, reduced swim pressure due to self-trapping of the particles \cite{sol15b} (and adverse external forces) and the non-monotonic density-dependence of the pressure \cite{yan14}. These phenomena, observed even if hydrodynamic interactions between the solvent and the active particles are neglected, are caused by active self-interactions of active particles contained in the tensor $\mathcal{I}_2$. Furthermore, our expression for the pressure shows manifestly the general lack of equivalence between the local pressure, defined via momentum exchange, and the mechanical pressure exerted on a wall for active systems, as previously argued \cite{spe16}. Our theory clarifies the role of activity as responsible for this general violation of (local) momentum conservation, putting in evidence the regimes in which such violation can be neglected and the stress tensor actually counterbalances the force exerted on a wall. \\
A straightforward extension of our theory would be the introduction of an angle-dependent inter-particles interaction able to enrich collective phenomena with aligning effects. Furthermore, it would be interesting to implement the effect of hydrodynamic interaction in our theory, in order to elucidate its role for the collective dynamics and to eventually derive a comprehensive dynamical density functional theory for a ``wet" active fluid. A first attempt would the introduction of a space dependent active velocity directly from the microscopic Langevin equation. 

\subsubsection*{Acknowledgment}
We acknowledge funding by Deutsche Forschungsgemeinschaft ({DFG}) via SPP 1726/1. S.S. acknowledges funding by International Max Planck Research Schools ({IMPRS}).

\appendix
\section{Multiple time scale analysis} \label{app}
In this section we sketch the systematic expansion in the high-friction limit by which we obtain the contributions to zero order \eqref{rho0} and to first order \eqref{rho1}, as reported in the main text. We rewrite the microscopic FPE \eqref{FPE} with the help of a multiple-time-scale analysis, introducing the small parameter $\epsilon$ to label the expansion of the FP density (as in the main text $\tilde{\gamma}=\epsilon \gamma$). The derivation follows that for a passive Langevin system \cite{boc97} and extends it to active systems, with slight manipulations. An expansion of the $N$-particle PDF is obtained comparing the terms in the FPE order by order. For this purpose,  the three time scales defined in the main text are introduced on the $\rm LHS$ of the FPE through the chain rule relation for the times, 
\begin{align}\label{eq:chain rule}
\frac{\partial}{\partial t}=\varepsilon^{-1}\frac{\partial}{\partial\tau_0}+\frac{\partial}{\partial \tau_1}+\varepsilon\frac{\partial}{\partial\tau_2}
\end{align}
Since the active force is proportional to $\gamma$ via the definition $F^{\text{A}}=\gamma v_0$, the zero order or fast time scale, corresponding to the terms proportional to $\gamma^{-1}$, is given by
\begin{align}\label{fast}
\frac{\partial \rho^{\left(0\right)}}{\partial\tau_0}=\mathcal{L}_{\text{FP}}\rho^{\left(0\right)}
\end{align}
where we defined the operator
\begin{align}\label{zeroroder}
\mathcal{L}_{\text{FP}}=\tilde{\gamma}\frac{\partial}{\partial v}\left(v-v_0n_i+T\frac{\partial}{\partial v}\right)
\end{align}
After a transient, \eqref{fast} describes the fast time scale $\tau_0$ a relaxation to $\rho^{\left(0\right)}=\Phi\left(x,\theta,\tau_1,\tau_2\right)\mathcal{W}$, the solution of $\mathcal{L}_{\text{FP}}\rho^{\left(0\right)}=0$, namely,
\begin{align}\label{gauss}
\mathcal{W}=\frac{1}{2\pi T}\exp\left(-\frac{\left(v-v_0n_i\right)^2}{2T}\right).
\end{align}
To this order, the local swim velocity is the only contribution to the hydrodynamic flow velocity. Note that \eqref{gauss} is for a $D=2$ system, but an extension to higher dimensions is straightforward. \\\\
First-order corrections can be found comparing the terms proportional to $\epsilon$ in the FPE, and using that the operator $\mathcal{L}_{\text{FP}}$ acts only on $\rho^{(1)}$. For simplicity we denote by $\mathcal{F}$ all forces appearing in the FPE but the active one , 
\begin{align}\label{firstorder}
\frac{\partial\rho^{\left(1\right)}}{\partial\tau_0}+\frac{\partial \rho^{\left(0\right)}}{\partial \tau_1}+v\frac{\partial \rho^{\left(0\right)}}{\partial x}+\mathcal{F}\frac{\partial \rho^{\left(0\right)}}{\partial v}=\mathcal{L}_{\text{FP}}\rho^{\left(1\right)}
\end{align}
On the fast timescale $\tau_0$, the solution $\rho^{(1)}$ relaxes to the solution of
\begin{align}\label{firstorder1}
\frac{\partial \rho^{\left(0\right)}}{\partial \tau_1}+v\frac{\partial \rho^{\left(0\right)}}{\partial x}+\mathcal{F}\frac{\partial \rho^{\left(0\right)}}{\partial v}=\mathcal{L}_{\text{FP}}\rho^{\left(1\right)}.
\end{align}
An integration over velocities eliminates the still unknown dependence of $\rho^{(1)}$ and leads to a solvability equation for $\Phi$
\begin{align}\label{Solv}
\frac{\partial\Phi}{\partial \tau_1}+v_0n_i\frac{\partial\Phi}{\partial x}=0.
\end{align}
In the derivation for passive systems \cite{boc97}, the solvability condition imposes $\partial_{\tau_1} \Phi=0$, while from \eqref{Solv} one sees that activity implies a dependence on $\tau_1$. In other words, for passive systems the intermediate time $\tau_1$ is only a formal expedient, here it is physically meaningful.
Introducing \eqref{Solv} in \eqref{firstorder1} and noting that
\begin{align}
\mathcal{L}_{\text{FP}}\left[\left(v-v_0n_i\right)\mathcal{W}\right]=-\tilde{\gamma}\left(v-v_0n_i\right)\mathcal{W}
\end{align}
the expression for $\rho^{(1)}$ can be obtained,
\begin{align}
\rho^{(1)}=-\frac{1}{\tilde{\gamma}}\left(v-v_0n_i\right)\left(\frac{\partial\Phi}{\partial x}-\frac{\mathcal{F}\left(x\right)}{T}\Phi\right)\mathcal{W}+\frac{1}{\tilde{\gamma}}\Psi\mathcal{W}.
\end{align}
where $\Psi\left(x,\theta,\tau_1,\tau_2\right)$ is still unknown. \\
For convenience, we present here the dynamical equations for the kinetic tensor $\sigma_{\rm kin}$ and the correlation tensor $\mathcal{C}_{nv}$ between orientation and velocity, $\mathcal{C}_{nv}$, which we use to obtain the high-friction limit of the fluid momentum equation. They read, respectively,
\begin{align}\label{sigmakin_eq}
&\frac 12 \partial_t\sigma_{\text{kin}}(r,t) =-\nabla_{r}\cdot\left\langle \sum_{i=1}^{N}v_{i}\left(v_{i}-u\right)^{2}\delta\left(x_{i}-r\right)\right\rangle \nonumber \\+&\gamma T\rho(r,t)+\left\langle \sum_{i=1}^{N}\left(-\gamma v_{i}+F^{\rm A}n_{i}\right)\left(v_{i}-u\right)\delta\left(x_{i}-r\right)\right\rangle \nonumber\\&\quad+\left\langle \sum_{i=1}^{N}\left(F_{i}^{\rm ext}+F_{i}^{\rm int}\right)\left(v_{i}-u\right)\delta\left(x_{i}-r\right)\right\rangle 
\end{align}

\begin{align}\label{cnv_eq}
&\partial_t\left(\rho\left(r,t\right)\mathcal{C}_{nv}\left(r,t\right)\right)=-\nabla_{r}\cdot \mean{ \sum_{i=1}^{N}n_{i}v_{i}v_{i}}_r \nonumber \\ -&\left(\gamma+D_{r}\right)\rho\left(r,t\right)\mathcal{C}_{nv}\left(r,t\right)+\gamma v_{0}\rho\left(r,t\right)\mathcal{Q}\left(r,t\right)\nonumber \\&\quad\qquad\qquad+\left(\mathcal{I}_1+\mathcal{I}_2\right)\rho\left(r,t\right).\phantom{\Bigg()}
\end{align}

\bibliography{activeHD}
\bibliographystyle{apsrev4-1}

\end{document}